\documentclass[aps,reprint,small,bf,hang,showpacs,showkeys,twocolumn]{revtex4-1}

\usepackage{times}
\usepackage[usenames,dvipsnames]{color}
\usepackage{dcolumn}
\usepackage{bm}
\usepackage{amssymb,amsmath}
\usepackage{graphicx}
\usepackage{subfig}
\usepackage{wasysym}

\newcommand{\degree}{\ensuremath{^\circ}}
\newcommand{\Varepsilon}{\text{\Large$\varepsilon$}}

\begin{document}

\title{Phase Transition in a Model of Y-Shaped Molecules}
\author{D. P. Ruth$^{1}$, R. Toral$^{2}$, D. Holz$^{3}$, J. M.
Rickman$^{1,4}$ and J. D. Gunton$^{1}$}

\affiliation{Department of Physics$^{1}$,\\ Lehigh University,
Bethlehem, PA
18015,\\
IFISC$^{2}$ (Instituto de F{\'\i}sica Interdisciplinar y Sistemas
Complejos, CSIC-UIB),
Palma de Mallorca, Spain,\\
Department of Physics$^{3}$,\\Drew University,
Madison, NJ 07940\\
Department of Materials Science$^{4}$,\\
Lehigh University, Bethlehem, Pa 18015
}


\begin{abstract}
In recent years the statistical mechanics of
non-spherical molecules, such as polypeptide chains and protein
molecules, has garnered considerable attention as their phase 
behavior has important scientific and health implications. 
One example is provided by immunoglobulin, which has a ``Y''-shape.
In this
work, we determine the phase diagram of Y-shaped molecules on a
hexagonal lattice through Monte Carlo Grand Canonical ensemble
simulation,
using histogram reweighting, multicanonical sampling, and finite-size
scaling. 
We show that (as expected) this
model is a member of the Ising universality class. For low
temperatures, we implemented multicanonical
sampling to induce faster phase transitions in the
simulation. By studying several system sizes, we use finite-size
scaling to
determine the two phase coexistence curve, including the bulk
critical temperature, 
critical chemical potential, and critical density.
\end{abstract}

\maketitle

\section{Introduction}
Our immune system is our primary defense against pathogenic organisms
and  malignant cells.  A major component of this system 
is the class of proteins known collectively as immunoglobulin G
(IgG). These are multidomain proteins that are particularly important
due to their ability to bind to antigens with remarkable specificity.
Antibodies have many practical applications due to this unusual
specificity: this includes a large number of diagnostic applications
\textit{in vitro}, such as immunofluorescence, western blotting,
and enzyme-linked immunosorbent assay analysis\cite{Perchiacca2012}.
Antibodies have also become of intense interest in the pharmaceutical
world, due to their efficacy as therapeutic molecules. This is
evidenced by the large number of antibodies that are either approved,
or in clinical trials for treating human disorders such as cancer,
rheumatoid arthritis, osteoporosis, and asthma\cite{Maggon2007,
Reichert2008, Reichert2005}. In such situations, highly concentrated
antibody solutions are required in order to have therapeutic effect.
However, such high concentration solutions present important
stability and delivery challenges, such as aggregation and large
solution viscosity. A large value of the viscosity occurs for IgG2,
for example, at concentrations of 150 mg/ml, apparently due to a
transition of the solution to a gel\cite{Cheng2013}. Therefore, understanding
the condensation of proteins in solutions of highly concentrated
immunoglobulin is of importance in pharmaceutical applications.\\
\\
 IgG molecules have a characteristic Y-shape, with two distal $F_{ab}$
arms that bind selectively to particular antigens. All of them have
the same size and Y-conformation, due to their common genetic basis.
However, they differ in their specific sequence of amino acids in the
variable domains of the $F_{ab}$ arms. As noted by Wang \textit{et al.}
\cite{Wang2013}, the variability in the amino acid sequences
can produce a large increase in the overall attractive interactions
between neighboring IgG molecules. As a consequence, these
interactions can cause a variety of phase transitions, including
reversible aggregation, liquid-liquid phase separation,
crystallization, and gelation, which occur in a wide variety of
proteins\cite{Gunton2007,Dumetz2008,Vekilov}.  Although
IgGs are typically quite soluble at physiological conditions,
sometimes they can become insoluble. In fact, recent studies of
protein condensation have been published both for  recombinant
pharmaceutical IgGs and monoclonal
IgGs\cite{Ahamed2007,Wang2011,Trilisky2011,Mason2011,Lewus2011,Nishi2010,Chen2010,Wang2012}.
A detailed discussion of the importance of IgGs in physiological and
pharmaceutical situations is given by Wang \textit{et
al.}\cite{Wang2013}, as well as Nezlin\cite{Nezlin2010}. Wang
\textit{et al.} stress that systematic studies of the phase behavior
and phase diagrams of IgG solutions are crucial for the understanding
of the pathological condensation in humans, as well as the stability
of antibody drug formulations. \\ 
\\
Of particular interest to this paper are the experimental studies of
liquid-liquid phase transitions  by Benedek {\it et
al} \cite{Wang2013,Wang2012,Wang2011}. They have reported several
important results in a recent study of liquid-liquid phase separation
in eight human myeloma IgGs and two recombinant pharmaceutical human
IgGs. The first thing to note is that all liquid-liquid coexistence
curves have quite similar shapes.  These curves differ from those
found for quasi-spherical proteins in that they are broad and
asymmetric and have relatively small critical concentrations.  The
similarity of their shapes is presumably due to their common Y-shape.
The critical temperatures of these liquid-liquid phase separations
vary from one IgG to another, due to the variability in the amino
acid sequences that leads to different net attractive protein-protein
interactions. These transitions are also metastable and are preempted
by the stable fluid-solid(freezing) transition, with the transition
temperatures typically in the range of $-20\degree$C to
$-30\degree$C. The study of such phase transitions is a relatively
new, but rapidly emerging research area.  It is therefore of interest
to examine the effects of such an unusual architecture on model
studies of phase transitions. One such study has already been carried
out on a model of IgG\cite{Li2008}. \\ 
\\
In this paper we present an initial study of Y-shaped molecules by
focusing on such molecules on two-dimensional lattices. Although this
model might lack some important features of the phase transitions
undergone by IgG in three dimensions (for example, it does not yield
any notable asymmetry in the phase diagram), it does provide an
example of a phase transition of a molecule with an unusual
architecture. In addition, there are examples of IgG on two
dimensional surfaces.  For example, human IgG has been adsorbed to
biomaterial surfaces, which can enhance long-term macrophage adhesion
\textit{in vitro}\cite{Jenney2000}. (However , in that case the Y molecules are not
aligned only in the plane, but have some alignment perpendicular to the plane, in contrast
to our model.)  Due to the geometry of the
molecule, we choose to use hexagonal lattices on which the
Y-molecules can be naturally placed. To study the possible phase
transition of our model, we use grand-canonical Monte Carlo
simulations along with the usual histogram reweighting  and
multicanonical biasing methods. The outline of our paper is as
follows:  In Section II we define our model and the method by which
we study it.  In Section III we present our results, namely
that this model belongs to the Ising universality class.  This 
particular result was expected {\it a priori} since the order parameter is a
scalar. Using finite-size scaling (FSS) methods, we also obtain the
liquid-liquid phase separation curve, including the location of the
critical point in the thermodynamic limit. We also provide results
for the finite-size system critical point, using a standard scaling
analysis. In Section IV we present a brief conclusion.

\section{Model and Simulation}
\begin{figure}
\centering
\includegraphics[width=0.55\linewidth]{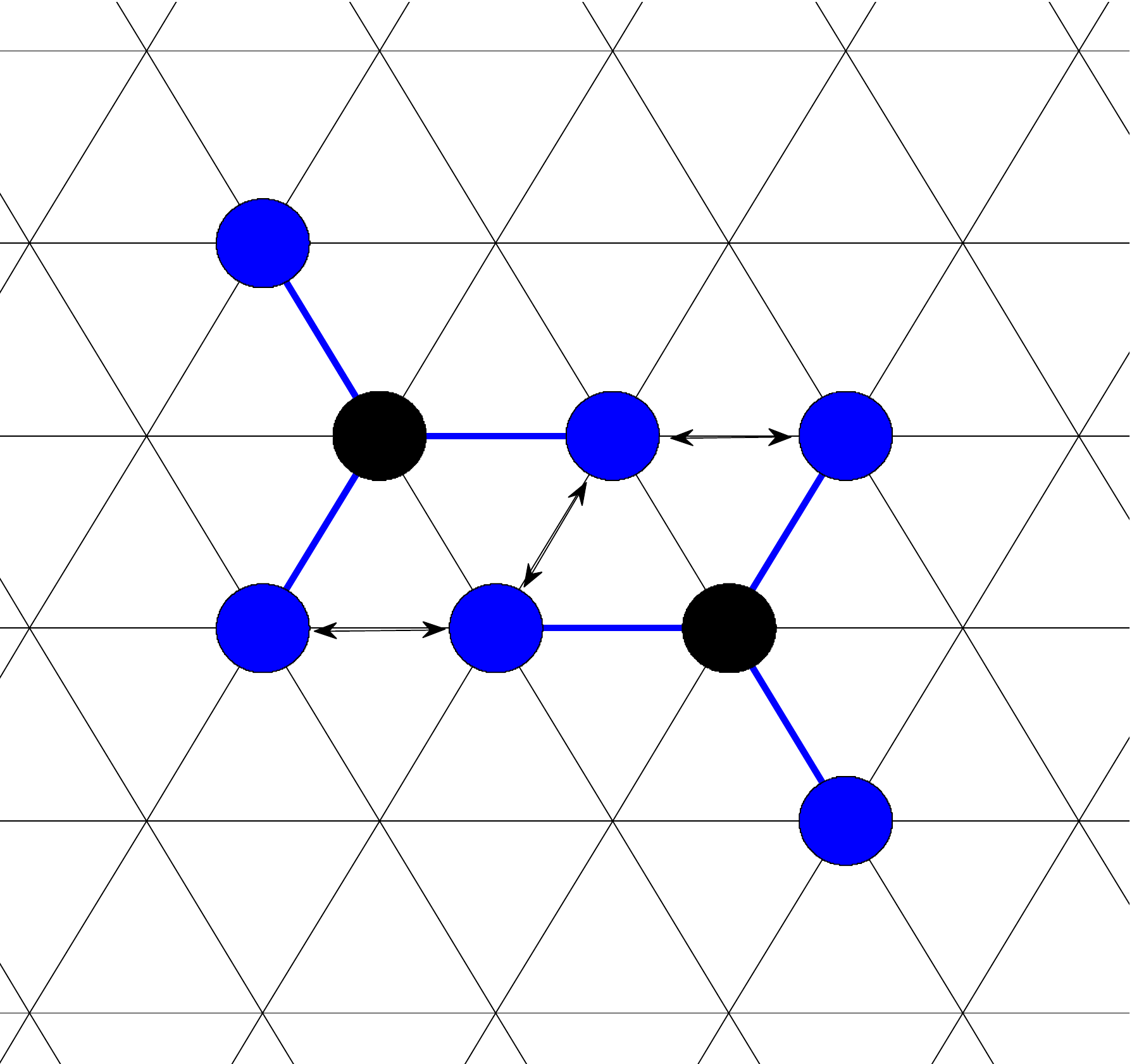}
\caption{Each center ({\color[RGB]{0,0,0}\CIRCLE}) and arm
({\color[RGB]{0,0,255}\CIRCLE}) 
occupy one lattice site, and each arm is physically bonded to the
center({\color[RGB]{0,0,255}\textbf{---}}). 
The interactions between arms and their nearest neighbors are denoted
by $\longleftrightarrow$.}
\label{fig:Model}
\end{figure}

Our model of the Y-shaped molecule is implemented within the
environment of a two-dimensional hexagonal lattice with 
side length $L$
containing $L^2$ sites. Periodic boundary conditions are employed.  
Each of the $N$
molecules occupies 4 lattice sites; the four include one central site
and 3 rigid {\sl distal} arms. This leads to a maximum number density
of the system $\rho$ of $0.25$. The interactions 
included in this study are those between distal arms, and their
respective nearest-neighbor distal arms, as shown in 
Fig.\ref{fig:Model}, each with interaction strength of $J$.
For simplicity, there are no center-to-center, center-to-distal arm, 
or any lattice binding energies. The total energy of the system $U$
is then the summation of all 
distal arm-distal arm interactions.
We execute grand-canonical Monte Carlo (GCMC) simulations. To analyze
the data and obtain the phase diagram, we use the Bruce-Wilding 
finite-size scaling (FSS)
techniques\cite{Bruce,Wilding}, 
along with histogram reweighting and multicanonical sampling
methods\cite{Toral-Colet:2014}, to compile the phase diagram of this
system.
The fact that the order parameter for this model (defined below) is a
scalar suggests that this Y-molecule model belongs to the 
Ising universality class, as we show in Section III.  

Assuming that our
model belongs to the Ising 
universality class, the critical point of our system can be
determined by matching the probability density function (PDF)
 of the ordering operator $M$ of our system with the universal
distribution of the two-dimensional Ising class. 
The order parameter $M$ for the fluid is given by\cite{Bruce,Wilding}
\begin{equation}
\label{eq:OrderParam}
M = \frac{1}{1-sr}[\rho - su],
\end{equation}
 where $u=U/N$ is the energy density, and $s$ and $r$ are system specific
parameters to be determined later. Similar to the order parameter
$M$, the energy-like parameter $\Varepsilon$ is given by 
\begin{equation}
\label{eq:EnergyParam}
\Varepsilon = \frac{1}{1-sr}[u - r\rho].
\end{equation}
The Ising universality class has two relevant scaling fields, namely 
$h$, the ordering scaling field,  and $\tau$, the 
thermal scaling field. For fluids in this universality class, $\tau$
and $h$ are defined as 
\begin{equation}
\label{eq:ScalingFields}
\tau = \omega_c - \omega + s(\mu-\mu_c),  h = \mu-\mu_c + r(\omega_c
- \omega),
\end{equation}
where $\omega = J/kT$, $\mu$ is the reduced chemical potential
in units of $kT$, and the subscript 
$c$ denotes the critical point. The parameters $r$ and $s$ determine
the degree of mixing in the relative scaling fields as well as 
$M$ and \Varepsilon. 

During a simulation in a system size of side $L$, at fixed values of
$\mu$ and $\omega$, we
record the molecule number density $\rho$ and the energy density $u$,
from which we determine the joint probability density function
$P(\rho, u)$.  The joint PDF, $P(M,\Varepsilon)$, for the rescaled
variables, $M$ and $\Varepsilon$, 
is related to the joint distribution of density and energy such that 
$P(M,\Varepsilon) = (1-sr)P(\rho,u)$. We focus mostly on the order
parameter PDF $P(M)=\int d\Varepsilon P(M,\Varepsilon)$. At the
critical point, all members of the Ising universality class have the
same fixed point distribution function. In the simulations, this fixed point distribution 
and the PDF of our model are expressed as $\tilde{P}_{M}(x)$ and $P_{L}(M)$ respectively, 
where $x = \alpha^{-1}_{M}L^{\beta/\nu}(M-M_{c})$. 
  $\beta=1/8$ and $\nu=1$ are the critical exponents of the order
parameter and correlation length of the two-dimensional Ising class,
respectively. $\alpha_{M}^{-1}$ is a scaling parameter such that
$\tilde{P}_{M}(x)$ has unit variance. Therefore, the PDF $P_{L}(x)$ of 
our model must also match $\tilde{P}_{M}(x)$ at the fixed point. It is only for large $L$ that the numerically obtained 
PDF tends to the fixed point distribution $\tilde{P}_{M}(x)$.

\begin{figure}
\includegraphics[width=0.70\linewidth]{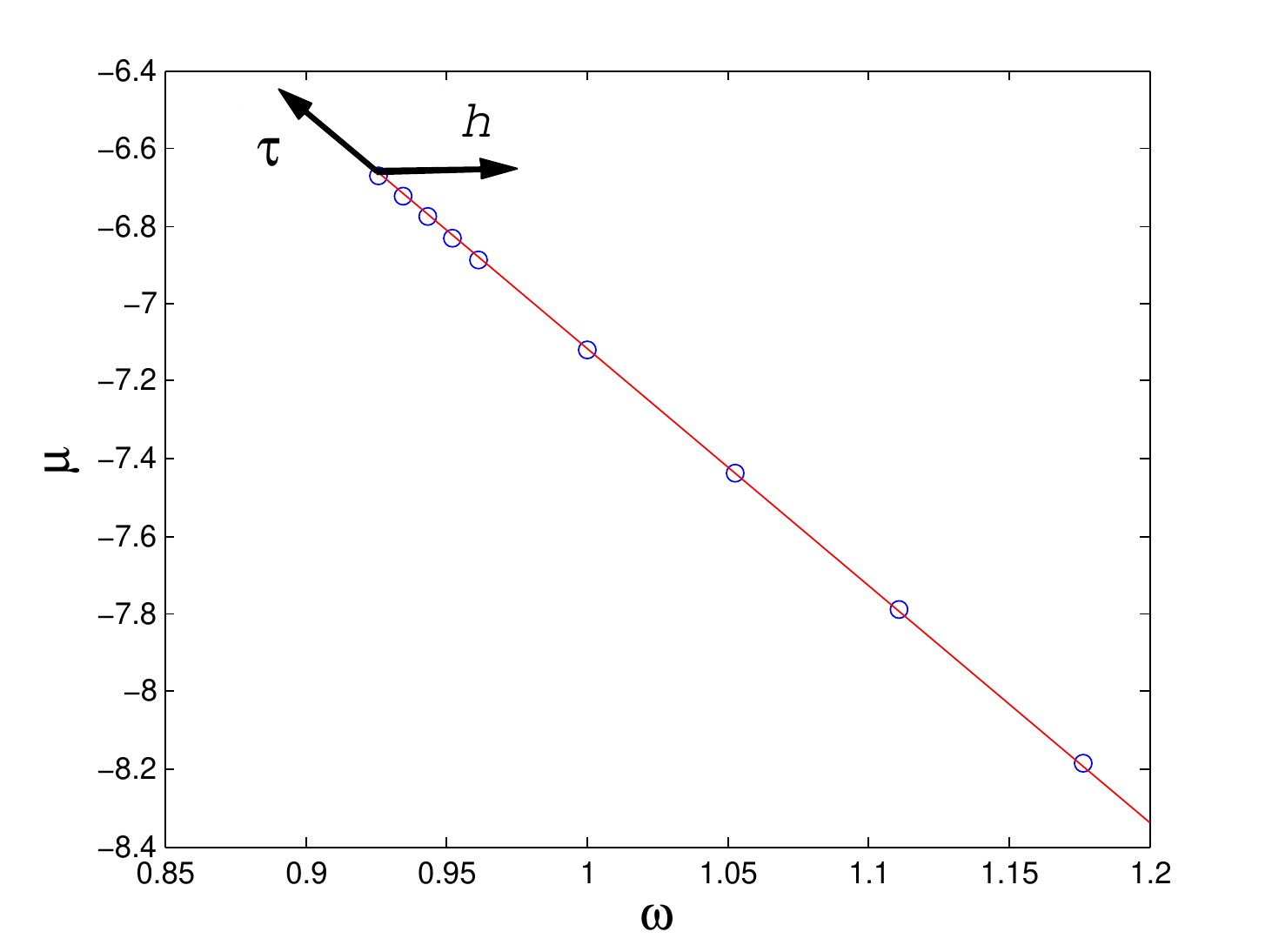}
\caption{Plot of reduced chemical potential $\mu$ versus the reduced 
interaction strength $\omega$. The
parameter $r$ is the slope of this curve at the critical temperature 
and is found to
 be $r= -6.111$.}
\label{fig:muvT}
\end{figure}

The fixed point function $\tilde{P}_{M}(x)$ for the two-dimensional
Ising model has been determined 
from previous work~\cite{Bruce}. For one system size, by varying $T$, $\mu$, and $s$, and
matching the numerically obtained $P_{L}(x)$ with $\tilde{P}_{M}(x)$,
we can determine the $T_{c}(L)$, $\mu_{c}(L)$, and $\rho_{c}(L)$ of our model at said system size.
By repeating this process for multiple system sizes, the bulk critical temperature($T_{c}$), 
chemical potential($\mu_{c}$), and density($\rho_{c}$) can be extrapolated. 
We obtain the parameter $r$ from the slope of the
$\mu$-$\omega$ 
coexistence line at criticality\cite{Wilding} as seen in Fig.\ref{fig:muvT}.

To avoid performing numerous simulations,  we use the standard method
of histogram reweighting. To analyze the data and obtain the phase
diagram, we use the Bruce-Wilding FSS techniques outlined
here\cite{Bruce, Wilding, Wilding1995, Lettieri}. In order to obtain
the critical parameters of the infinite system, we performed GCMC
simulations for systems with side lengths 30, 40, 50, and 60 with
periodic boundary conditions. The observables recorded during
the simulation were $u$ and $\rho$, from which, 
$P(\rho,u)$, $P(\rho)$, $P(M,\varepsilon )$, and others were
calculated. For each temperature, chemical potential, and system
size, the 
simulation ran for 5,000 - 6,000 Monte Carlo steps (MCS), and for 15,000 -
25,000  Monte Carlo steps 
for simulations implementing and not implementing biasing techniques
respectively, before recording the density and energy of the system.
Each MCS comprises $N$ attempts to change the system either by
a molecule translation, rotation, insertion, or removal. The changes
that the GCMC attempts to make to the system are defined as follows:
\begin{description}
	\item[Translation] Attempt to move the center of a randomly selected
molecule to one of the 6 nearest neighboring sites of the center. 
	\item[Rotation] Attempt to rotate a randomly selected molecule
either clockwise, or counter-clockwise about the molecule center.
	\item[Insertion] Attempt to place a molecule, with a random
orientation, on a randomly chosen lattice site.
	\item[Removal] Attempt to remove a randomly selected molecule.
\end{description}
This algorithm is ergodic. The density and energy of the system are
recorded a total of 150,000 - 250,000 times over the length of the
simulation, with 1,000 - 2,000 and 2,000 - 4,000 MCS in
between each recording for systems with and without the 
preweighting function, respectively. 

\section{Results}
\begin{figure}
\includegraphics[width=0.70\linewidth]{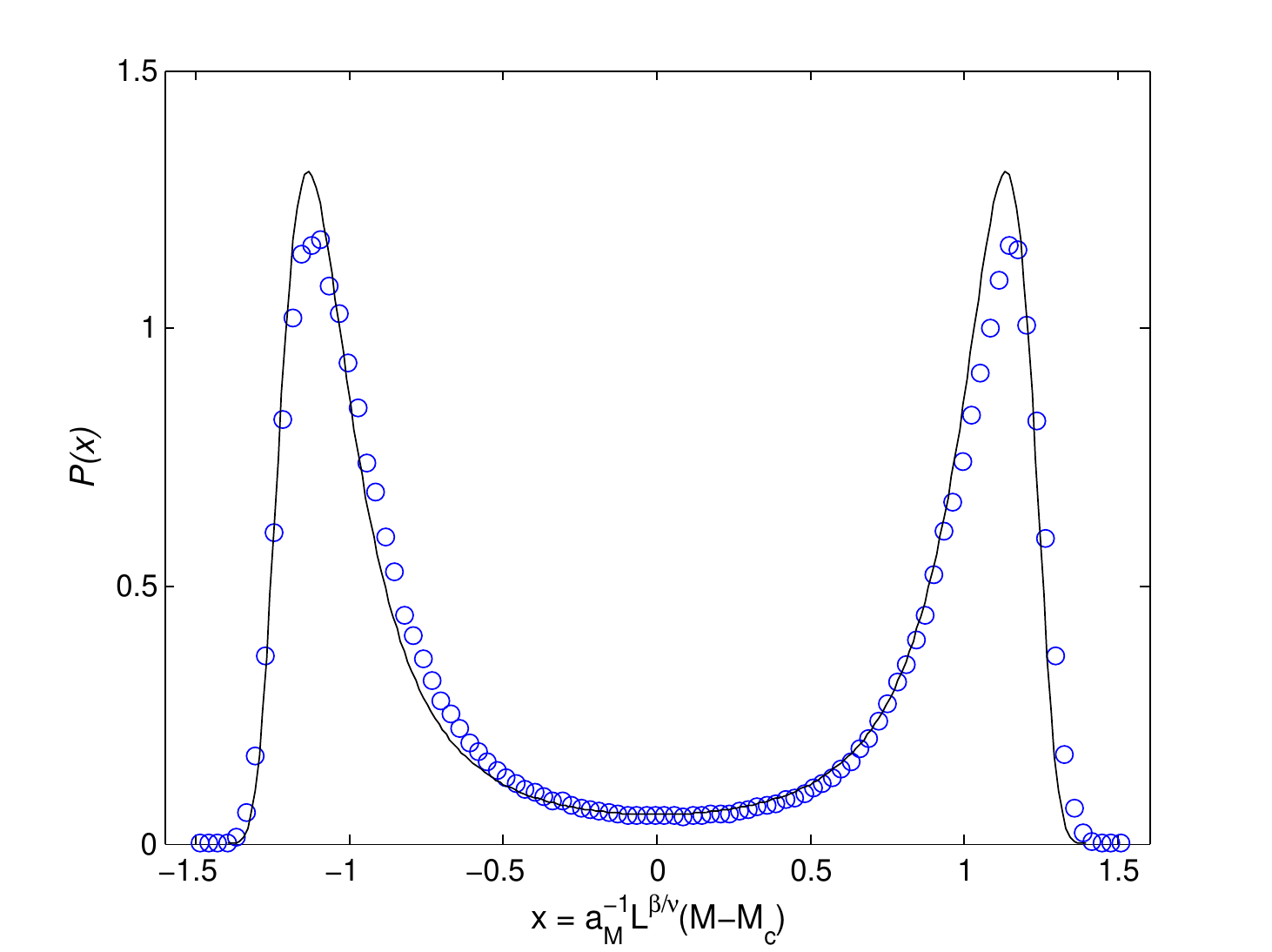}
\caption{$P_{L}(x)$ (${\color[RGB]{1,1,180}\circ}$) for $L =60$ and
the universal fixed point distribution $\tilde{P}_{M}(x)$(--) versus $x$. 
$T_{c}(L)$ and $\mu_{c}(L)$ were $T_{c}(L)=1.084$ and $\mu_{c}(L) =
-6.477$. Reproduced by permission of IOP Publishing, N. B. Wilding, and A. D. Bruce, ``Density fluctuations and field mixing in the critical fluid'' Phy.: Condens Matter \textbf{4}, 3087-3108 (1992). Copyright 1992 by IOP Publishing. All rights reserved\cite{Wilding}.}
\label{fig:UniversalOrder}
\end{figure}

\begin{figure}
\subfloat[Apparent $T_{c}(L)$ vs
$L^{-(\theta+1)/\nu}$]{\includegraphics[width=0.48\linewidth]{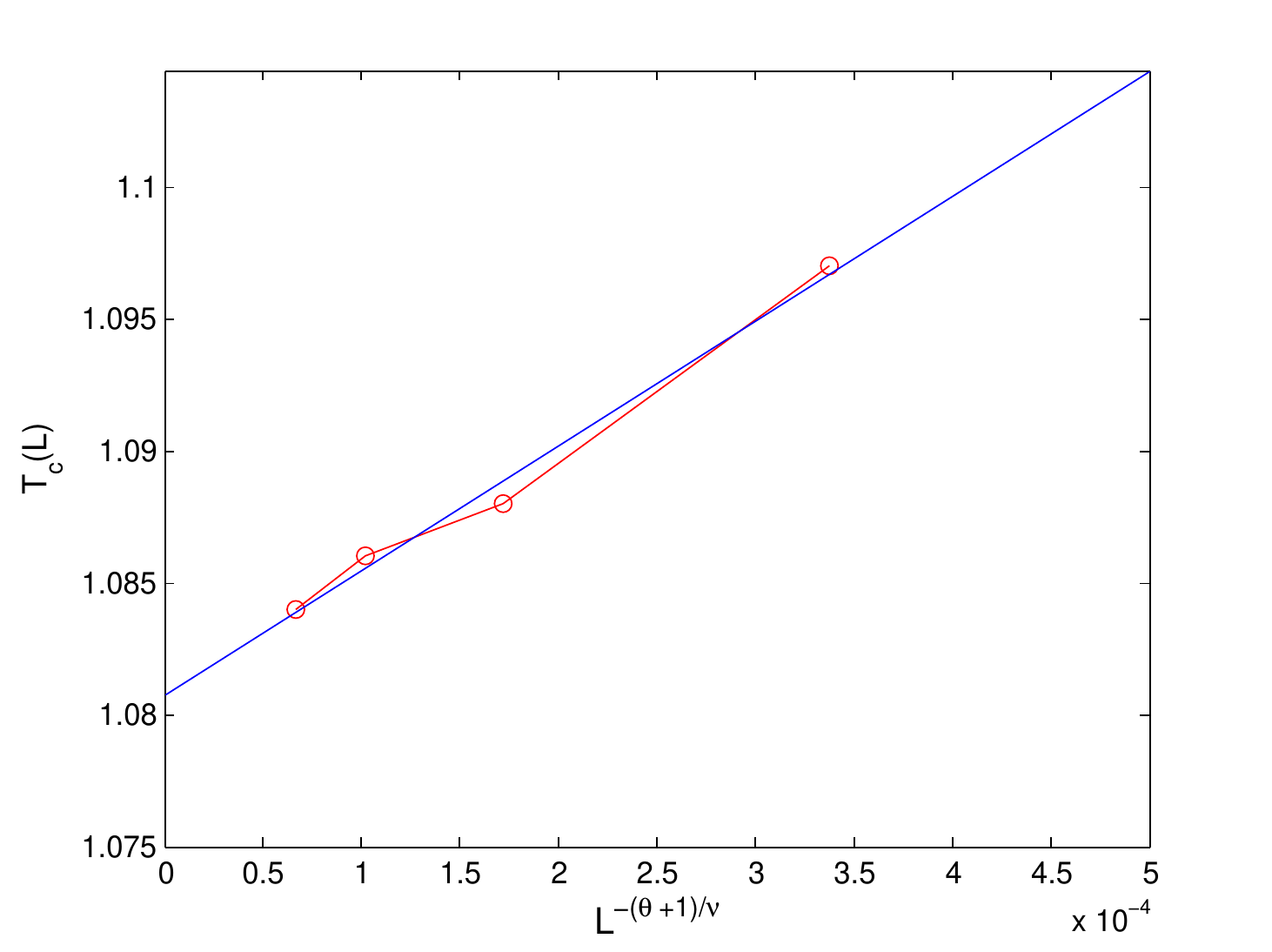}}
\label{fig:ApTvsL}
\quad
\subfloat[Apparent $\mu_{c}(L)$ vs
$L^{-(\theta+1)/\nu}$]{\includegraphics[width=0.48\linewidth]{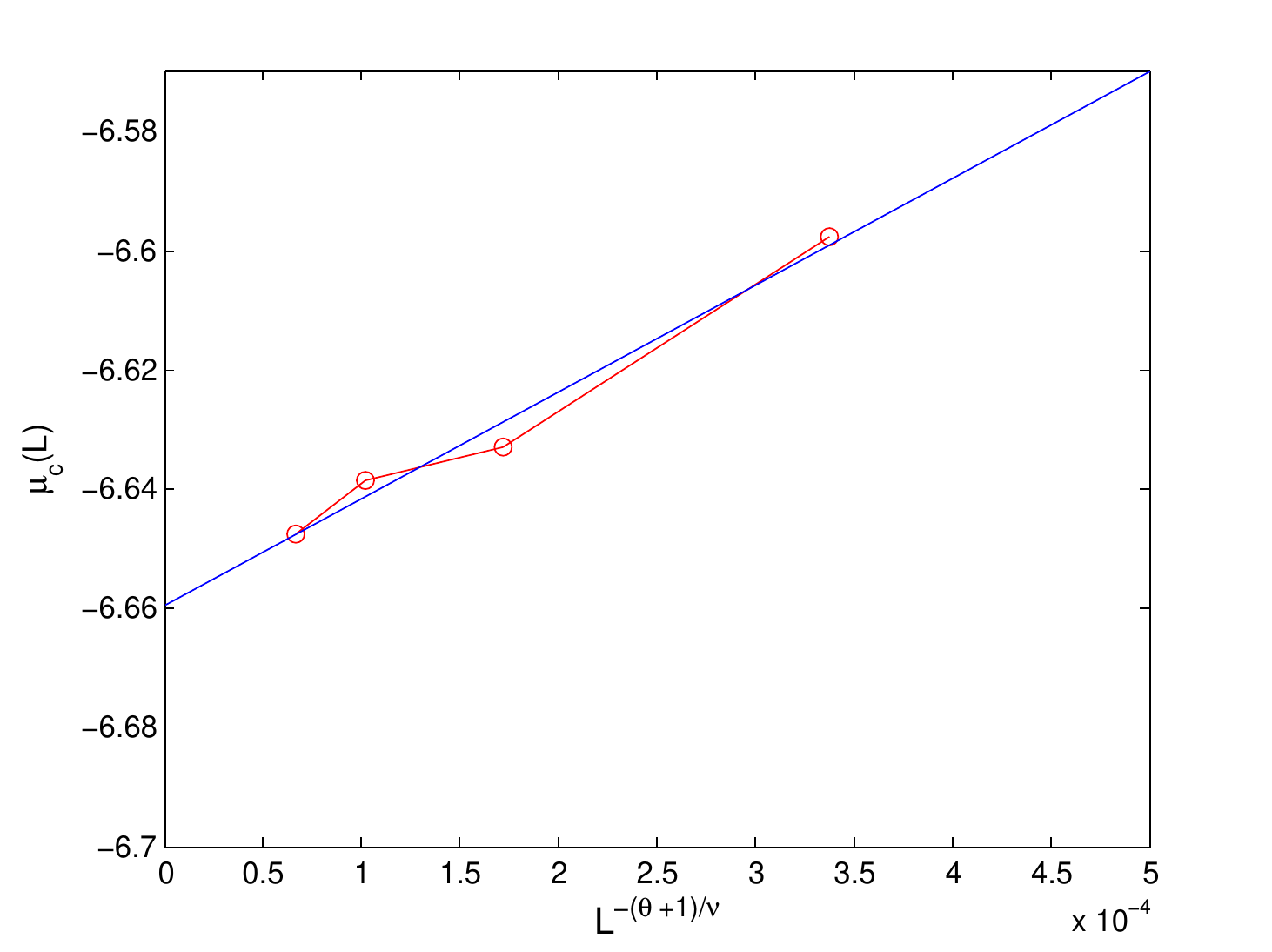}}
\label{fig:ApMUvsL}
\quad
\caption{a.) Plot of apparent bulk critical temperature determined
from matching $P_{L}(x)$ to $\tilde{P}_{M}(x)$. b.)  Plot of the apparent 
bulk critical chemical potential determined from matching $P_{L}(x)$
to $\tilde{P}_{M}(x)$. The extrapolated critical temperature and 
chemical potential are $T_{c}=1.081\pm0.002$ and $\mu_{c}=-6.66\pm0.01$.}
\label{fig:FSS}
\end{figure}

The method we use to find the bulk critical parameters is as follows.
First, we plot $P_{L}(x)$ 
for varying system sizes and 
find best fits to the universal fixed point $\tilde{P}_{M}(x)$ by
varying $T_{c}(L)$, $\mu_{c}(L)$, and $s$. A sample best fit of
$P_{L}(x)$ to this fixed 
point is shown in Fig.\ref{fig:UniversalOrder} and shows that, within
the accuracy of our study, our model belongs to the Ising
universality class. Next, we use the FSS
predictions\cite{Wilding1995} that 
\begin{equation}
T_{c} - T_{c}(L) \propto L^{-(\theta+1)/\nu}
\label{eq:TFSS}
\end{equation}
and 
\begin{equation}
\mu_{c} - \mu_{c}(L) \propto L^{-(\theta+1)/\nu}
\label{eq:muFSS}
\end{equation} 
 to determine the bulk values of the critical temperature and
critical chemical potential. In these equations, $T_{c}$ and
$\mu_{c}$ are the actual bulk critical temperature and chemical
potential, and  $T_{c}(L)$ and 
$\mu_{c}(L)$ are the apparent bulk critical temperature and chemical
potential determined from matching $P_{L}(x)$ to the 
fixed universal distribution $\tilde{P}_{M}(x)$. $\theta$ is a
correction to scaling exponent. We use the value $\theta = 1.35$, as 
calculated by Barma and Fisher\cite{Fisher1985}, which coincides with 
the
conjecture of Nienhuis\cite{Nienhuis} for the two-dimensional Ising
system.
 The next step is to record $T_{c}(L)$ 
and $\mu_{c}(L)$ for each system size, plot $T_{c}(L)$ and
$\mu_{c}(L)$ versus $L^{-(\theta+1)/\nu}$, and then 
extrapolate to the infinite system size for both $T_{c}(L)$ and
$\mu_{c}(L)$. We then record the 
extrapolated points as $T_{c}$ and $\mu_{c}$. The resulting graphs of
$T_{c}(L)$ and $\mu_{c}(L)$ versus $L^{-(\theta+1)/\nu}$ are given in
Fig.\ref{fig:FSS}. 

\begin{figure}
\centering
\includegraphics[width=0.70\linewidth]{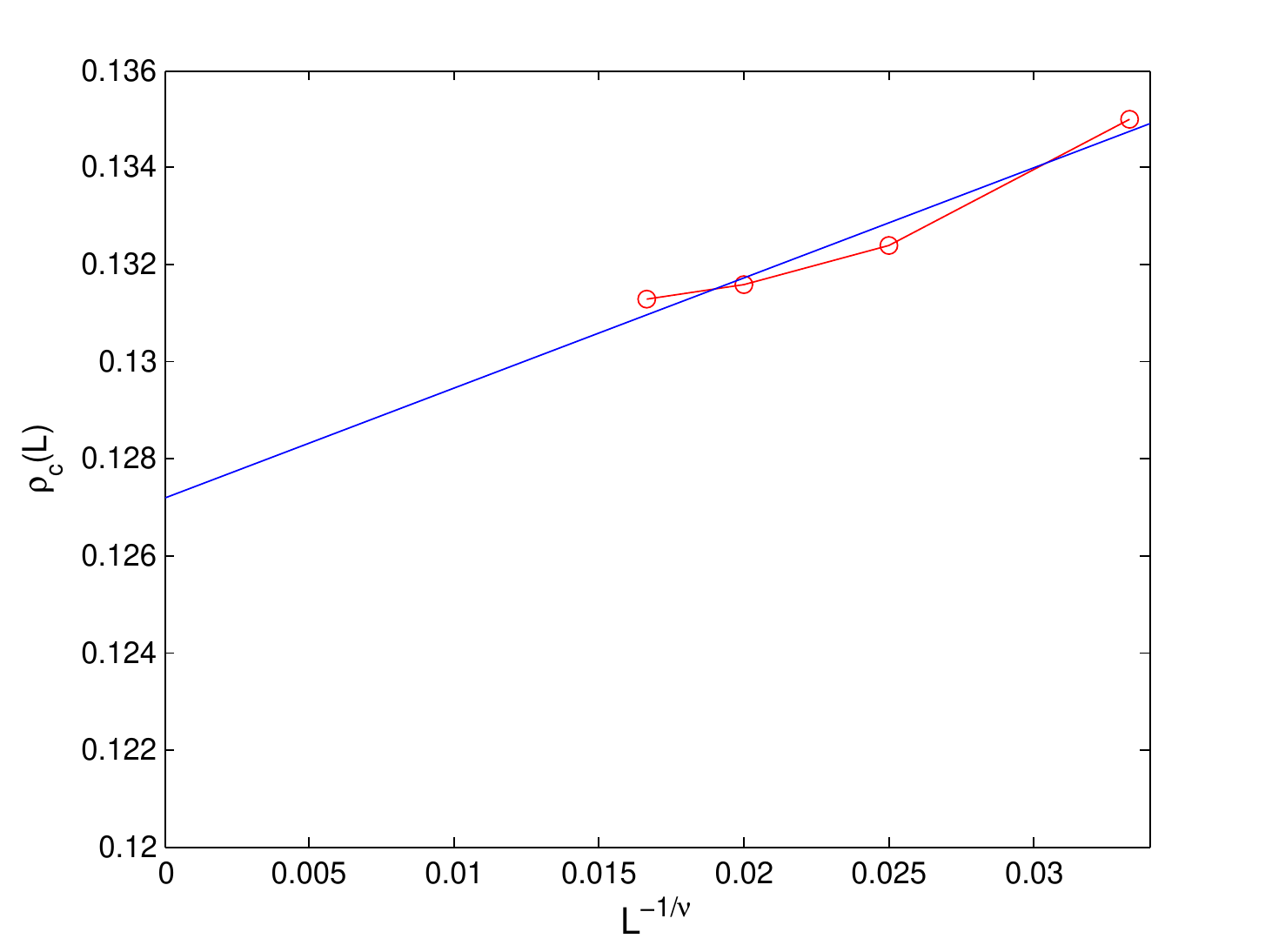}
\caption{Plot of the apparent $\rho_{c}(L)$ vs $L^{-1/\nu}$. The extrapolated
$\rho_{c}=0.127\pm0.002$.}
\label{fig:rhoFSS}
\end{figure}

FSS also predicts a similar correction to the critical density of the
model, namely
\begin{equation}
\rho_{c} - \rho_{c}(L) \propto L^{-(d-1/\nu)},
\label{eq:rhoFSS}
\end{equation}
where $\rho_{c}$ is the bulk critical density, $\rho_{c}(L)$ is the
apparent critical density of a system of size $L$ at 
$T_{c}$ and $\mu_{c}$, and $d = 2$ is the dimensionality of the
system. Similar to the cases of $T_{c}$ and $\mu_{c}$, 
$\rho_{c}$ is extrapolated from the plot in Fig.\ref{fig:rhoFSS} and
is  $\rho_{c}=0.127\pm0.002$.

\begin{figure}
\centering
\includegraphics[width=0.70\linewidth]{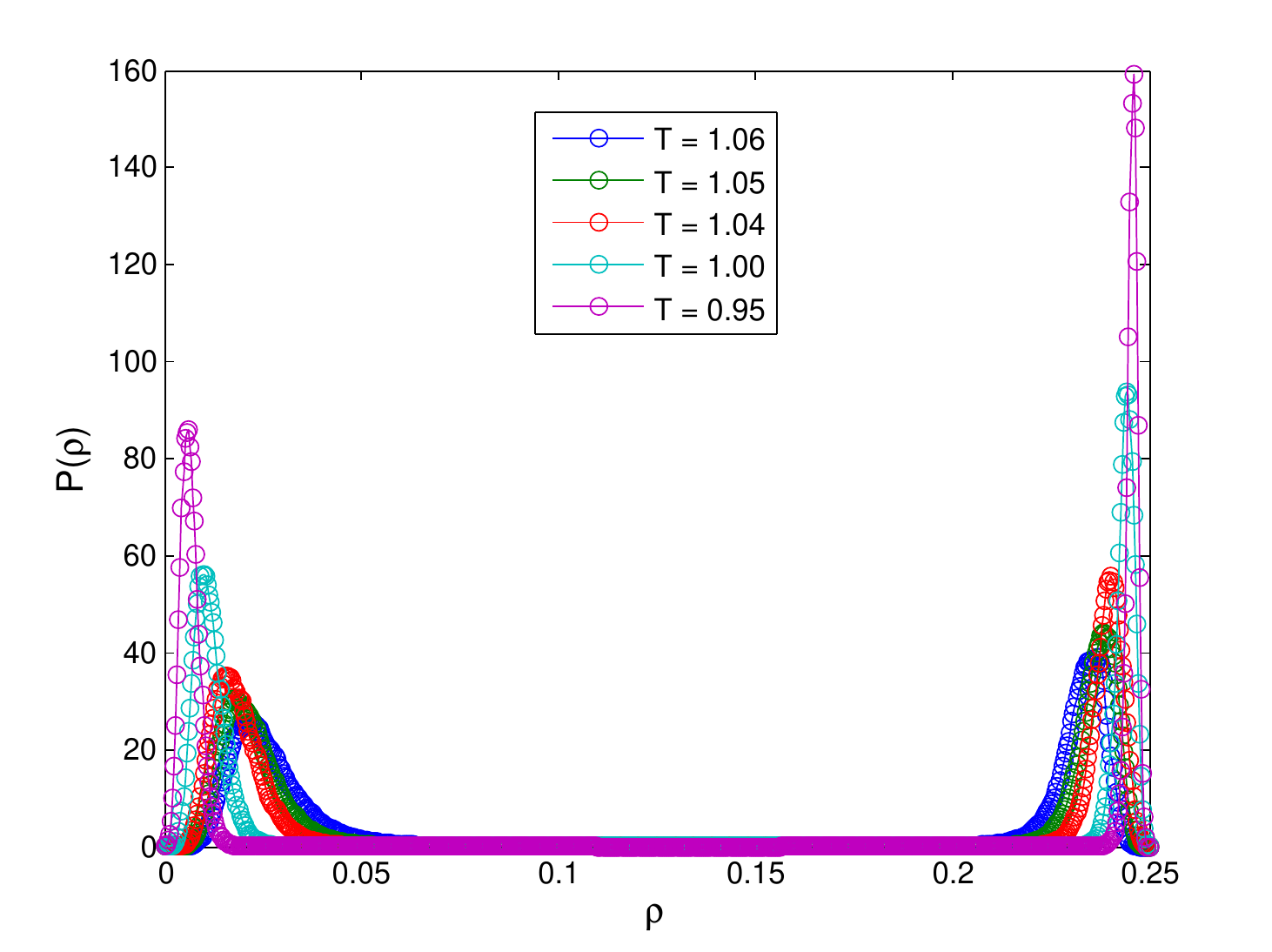}
\caption{Plot of estimated $P(\rho)$ versus $\rho$ for varying
temperatures at coexistence. All $P(\rho)$ were determined as
described in 
the text.}
\label{fig:Prho}
\end{figure}

The coexistence curve was determined through a series of GCMC
simulations at varying temperatures below the critical region,
implementing the histogram reweighting and multicanonical sampling
techniques discussed previously. Coexistence between two phases at a 
temperature $T$ is confirmed when the areas underneath the two peaks
in the density distribution $P(\rho)$ are equal. The peak 
densities are recorded and plotted on the phase diagram. Examples of
some density distributions for varying $T$ at coexistence are 
shown in Fig.\ref{fig:Prho}. 

\begin{figure}
\centering
\includegraphics[width=0.70\linewidth]{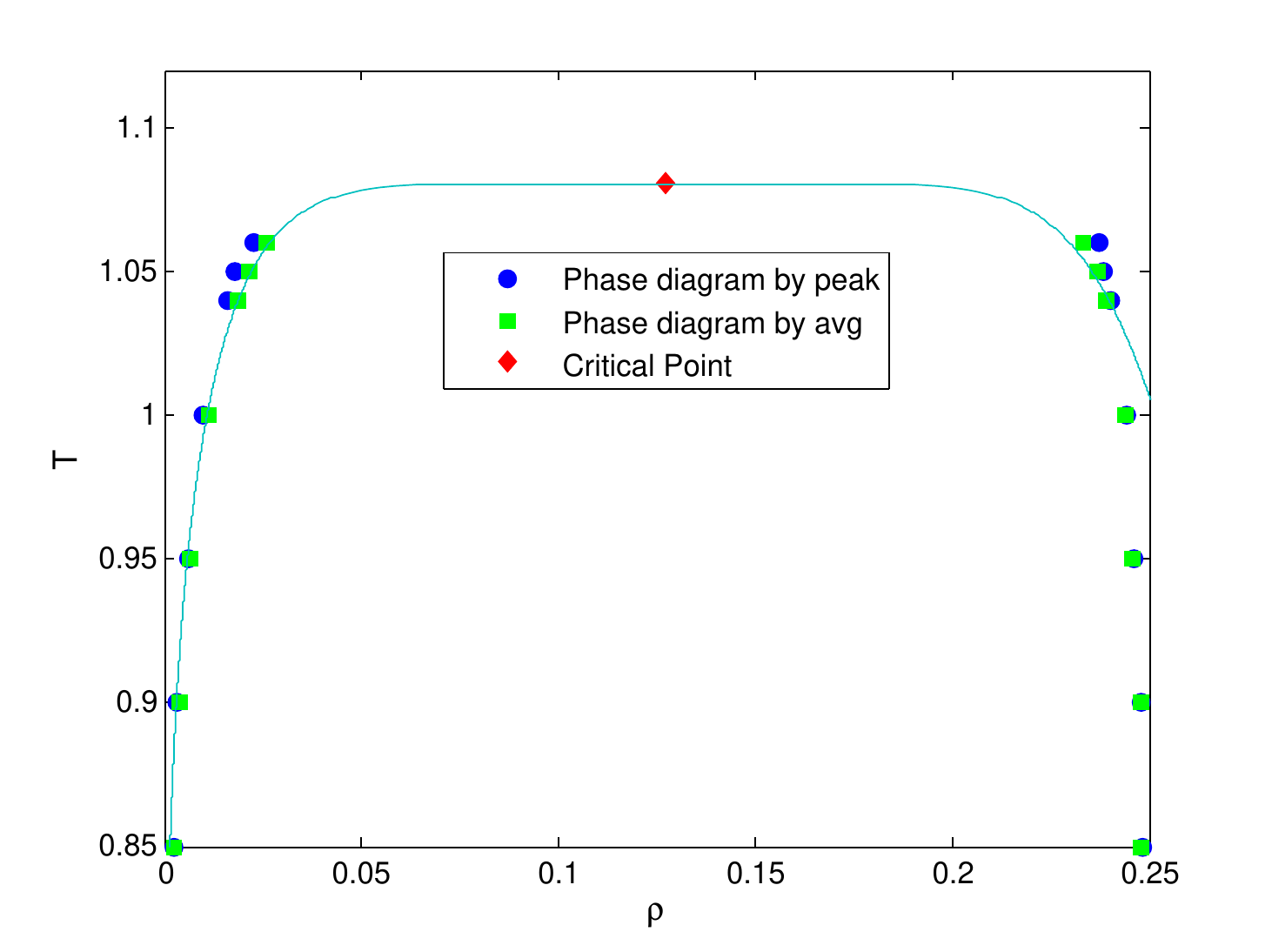}
\caption{The reduced temperature versus density, as obtained by the two
methods discussed in the text. The values obtained from the positions
of the maxima of the probability distribution functions 
(${\color[RGB]{0,0,255}\CIRCLE}$) and from the
average values (${\color[RGB]{0,255,0}\blacksquare}$)are plotted as
a function of temperature. Also shown is the best fit to data through
$T_{c}$ and $\rho_{c}$ of 
the form $\rho \pm 
\rho_{c} = a|T-T_{c}| \pm b|T-T_{c}|^{\beta}$ with $a = 0.05$ and $b
= 1.65$.}
\label{fig:PhaseDiagram}
\end{figure}

In temperature regions apart from the critical region, finite-size
effects are expected to be negligible since the correlation length is
much 
smaller than that of the size of the system. With this expectation,
the density peaks of $P(\rho)$ found in our finite systems at
different 
temperatures below the critical region will still mimic that of an
infinitely large system. A phase diagram of the infinite system is
constructed 
by determining the positions of the peaks in $P(\rho)$ for
sub-critical temperatures. As an additional check on this estimate of
the equilibrium densities, we also calculate the average density of
each the two phases present by using the PDFs to calculate these statistical averages. 
Our results are
shown in Fig.\ref{fig:PhaseDiagram}, where the FSS estimates of $T_{c}$ and
$\rho_{c}$ are also plotted.

We fit the points of our phase diagram to a power law of the
form\cite{Wilding1995} 
\begin{equation}
\label{eq:PowerLawFit}
\rho \pm \rho_{c} = a|T-T_{c}| \pm b|T-T_{c}|^{\beta}.
\end{equation}
 
\begin{figure}
\centering
\includegraphics[width=0.70\linewidth]{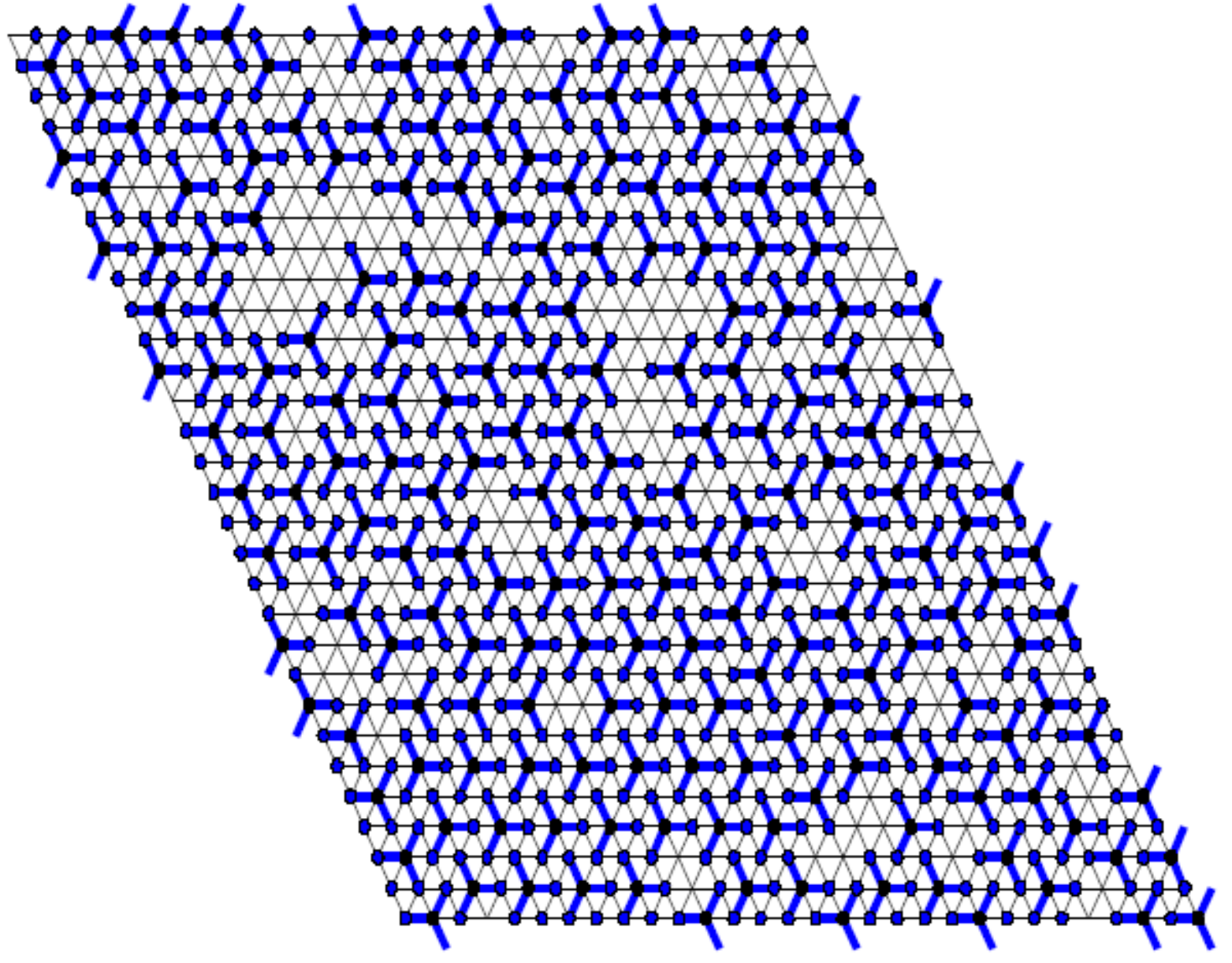}
\caption{Sample of system size $L=30$, at number density $\rho=0.2156$
and temperature $T= 1.08$.}
\label{fig:Lattice}
\end{figure}

This fit is also included in Fig.\ref{fig:PhaseDiagram}.
There is a small asymmetry evident on the high density side of the
phase diagram, but not as pronounced as the one found experimentally 
or numerically in 3D for IgG.
We show the critical parameters determined through the FSS method 
in Table \ref{Tab:Parameters}. We also show in Fig.\ref{fig:Lattice} 
a typical configuration of a
high density system in equilibrium.
\begin{table}
\begin{center}
\begin{tabular}{|l|l|l|l|l|}
\hline
$r$ & $s$ & $T_{c}$ & $\mu_{c}$ & $\rho_{c}$ \\
\hline
$-6.111$ & $\simeq 0.05$ & $1.081\pm0.002$ & $-6.66\pm0.01$ & $0.127\pm0.002$ \\ 
\hline
\end{tabular}
\caption{List of all critical point parameters}
\label{Tab:Parameters}
\end{center} 
\end{table}
\section{Conclusion}
We end this paper with a few comments. First, the major feature of
our work is that we have determined in detail using finite-size
scaling methods the phase diagram of a  molecule with unusual
architecture. We have chosen a very simple model for the interaction
between these molecules; it would be straightforward to include other
interactions in future work. We have shown that the model belongs to
the Ising universality class as one might expect, since its order
parameter is a scalar. It seems clear that future research will
increasingly deal with unusual molecular architectures. We note in
passing the fact that the molecule has a Y-shape has not resulted in
an asymmetry of the type found for IgG on the high
density side. There are several reasons for this.  First, our model
is two-dimensional.  Second, we   have not taken into account in the
model the large excluded volume effect that characterizes IgG, which
Bendek {\it et al.} \cite{Wang2013} have argued is responsible for its
asymmetry. Indeed, the three-dimensional model of IgG studied by Li
{\it et al.} \cite{Li2008} does have an excluded volume effect, and the phase
diagram shows a pronounced asymmetry similar to that observed in
experiment. It should also be noted that the reason our model has an
asymmetry is due to the absence of a particle-hole symmetry, such as 
that present in the Ising model.\\
\\
One direction for future research is to improve our model in order to
describe the absorption of IgG on surfaces\cite{Li2008}.  The 
inclusion of surfaces would require a somewhat more complex 
simulation algorithm, but would have the benefit of producing richer phase 
behavior.  Another research direction would involve a more chemistry based, 
coarse-grained model of
IgG, such as the one proposed by Voth's group \cite{Chaudri2013}, to
study not only its phase transitions, but its viscosity.  Such models 
are especially useful in probing the electrostatic interaction 
between antibodies and its impact on ordering.  Finally, we note that 
the rheological information (such as the viscosity) that could be obtained with a 
coarse-grained model is of considerable interest in the pharmaceutical world.
\section{Acknowledgement} This work was supported by a grant from the
G. Harold and Leila Y. Mathers Foundation. RT acknowledges financial
support from EU (FEDER) and the Spanish MINECO under Grant
INTENSE@COSYP (FIS2012-30634). DH acknowledges financial support from  NSF 
PHY-0849416 and PHY-1359195.

\bibliography{Ch4bib}

\end{document}